\documentclass[conference]{IEEEtran}
\IEEEoverridecommandlockouts

\usepackage{cite}
\usepackage{amsmath,amssymb,amsfonts}
\usepackage{algorithmic}
\usepackage{graphicx}
\usepackage{textcomp}
\usepackage{xcolor}
\def\BibTeX{{\rm B\kern-.05em{\sc i\kern-.025em b}\kern-.08em
    T\kern-.1667em\lower.7ex\hbox{E}\kern-.125emX}}
    
\begin{document}
\bstctlcite{IEEEtran:BSTcontrol}

\title{MoRE: A Mixture-of-Experts-Based Task-Adaptive End-to-End Network for Multimodal MRI Reconstruction}

\makeatletter
\newcommand{\linebreakand}{%
  \end{@IEEEauthorhalign}
  \hfill\mbox{}\par
  \mbox{}\hfill\begin{@IEEEauthorhalign}
}
\makeatother

\author{
\IEEEauthorblockN{1\textsuperscript{st} Yuyang Li}
\IEEEauthorblockA{\textit{School of Biomedical Engineering} \\
\textit{ShanghaiTech University}\\
Shanghai, China \\
liyuyang@shanghaitech.edu.cn}
\and
\IEEEauthorblockN{2\textsuperscript{nd} Yipin Deng}
\IEEEauthorblockA{\textit{School of Biomedical Engineering} \\
\textit{ShanghaiTech University}\\
Shanghai, China \\
dengyp2024@shanghaitech.edu.cn}
\and
\IEEEauthorblockN{3\textsuperscript{rd} Wenlei Shang}
\IEEEauthorblockA{\textit{School of Biomedical Engineering} \\
\textit{ShanghaiTech University}\\
Shanghai, China \\
shangwl2022@shanghaitech.edu.cn}
\linebreakand

\IEEEauthorblockN{4\textsuperscript{th} Juncen Wu}
\IEEEauthorblockA{\textit{School of Biomedical Engineering} \\
\textit{ShanghaiTech University}\\
Shanghai, China \\
wujc2024@shanghaitech.edu.cn}
\and

\IEEEauthorblockN{5\textsuperscript{th} Xin Bai}
\IEEEauthorblockA{\textit{School of Biomedical Engineering} \\
\textit{ShanghaiTech University}\\
Shanghai, China \\
baixin2023@shanghaitech.edu.cn}

\and
\IEEEauthorblockN{6\textsuperscript{th} Zijian Zhou}
\IEEEauthorblockA{\textit{School of Biomedical Engineering} \\
\textit{ShanghaiTech University}\\
Shanghai, China \\
zhouzj@shanghaitech.edu.cn}

\linebreakand

\IEEEauthorblockN{7\textsuperscript{th} Peng Hu}
\IEEEauthorblockA{\textit{School of Biomedical Engineering} \\
\textit{ShanghaiTech University}\\
Shanghai, China \\
hupeng@shanghaitech.edu.cn}
}

\maketitle

\begin{abstract}
Although accelerated MRI reconstruction has advanced rapidly through end-to-end learning, deploying a single unified network that generalizes across diverse anatomies and contrasts under constrained computational resources remains challenging. In this paper, we introduce MoRE, a sparsely activated mixture-of-experts (MoE) module integrated into an end-to-end variational network. MoRE couples a shared encoder with sample-wise, unsupervised routing to activate a minimal subset of expert decoders while strictly preserving physics-based data consistency. Evaluated on the fastMRI multi-coil brain and knee datasets under 8× undersampling, MoRE achieves highly stable SSIM and PSNR performance across multi-contrast datasets. Furthermore, t-SNE visualization of the routing embeddings reveals interpretable, modality-aware expert specialization. The sparse conditional computation mechanism ensures that the architectural overhead remains modest. These results demonstrate that MoE-style capacity scaling can significantly enhance general-purpose MRI reconstruction without requiring proportional increases in computational power.
\end{abstract}


\section{Introduction}
\label{sec:intro}

Magnetic resonance imaging (MRI) provides exceptional soft-tissue contrast, detailed anatomical visualization, and quantitative information, making it indispensable in modern clinical practice \cite{cardiacmri}. However, the lengthy acquisition required for fully sampled k-space measurements remains a major limitation. To reduce scan time, accelerated MRI techniques acquire undersampled measurements and recover missing information using methods such as parallel imaging \cite{HAMILTON201771}, compressed sensing, and deep learning \cite{8962951}. Based on these advances, end-to-end trained neural networks have demonstrated superior performance over classical reconstruction methods in many high-acceleration settings \cite{Muckley_2021}. Among them, End-to-End Variational Networks (E2EVarNet) \cite{sriram2020endtoendvariationalnetworksaccelerated} are particularly compelling, as they combine learned image priors with explicit physics-based data consistency, producing robust reconstruction performance and a more interpretable framework \cite{Muckley_2021,lyu2024stateoftheartcardiacmrireconstruction}.

Despite this progress, deploying deep reconstruction models in practice remains challenging due to the competing demands of cross-task generalization and computational efficiency. While conventional learning-based reconstruction models are often trained for specific anatomies, contrasts, or sampling settings, recent studies suggest that careful network design, training strategies, and dataset composition can enable a single model to generalize across tasks \cite{wang2025universallearningbasedmodelcardiac}. In addition, recent evidence indicates that training on heterogeneous data distributions can yield robustness comparable to, or even better than, training on a single distribution, whereas overfitting to a single dataset may degrade out-of-distribution reconstruction quality \cite{lin2024robustnessdeeplearningaccelerated}. Meanwhile, recent benchmarks and challenge studies suggest that higher-capacity networks often achieve stronger general-purpose performance \cite{Muckley_2021,lyu2024stateoftheartcardiacmrireconstruction,wang2025universallearningbasedmodelcardiac}. However, this improvement typically comes with substantial computational and memory demands, which may limit adoption in resource-constrained academic settings.

A promising direction is the mixture-of-experts (MoE) paradigm \cite{mu2026comprehensivesurveymixtureofexpertsalgorithms}, which has gained increasing attention as a scalable approach to increasing model capacity. Unlike conventional dense modules that activate all parameters for every input, MoE conditionally routes inputs to a sparsely selected subset of experts based on input features. This mechanism fosters expert specialization and can enable substantial parameter scaling without a proportional increase in per-sample computation. Moreover, MoE-induced sparsity has been reported to improve robustness in multitask learning \cite{gupta2022sparselyactivatedmixtureofexpertsrobust}. Although MoE has been widely adopted in natural language processing, recent studies have also extended it to medical imaging. Luo et al. \cite{Luo_Zhu_Zhang_Sun_2025} introduced mixture skip connections into U-Net for multitask segmentation, while another study combined a ViT \cite{dosovitskiy2021imageworth16x16words} with an MoE feedforward network to achieve spatial, patch-wise adaptive MRI denoising \cite{Deng_2025}. However, the use of MoE for physics-constrained, end-to-end MRI reconstruction remains underexplored.

We summarize our main contributions as follows:

\begin{itemize}
  \item We propose MoRE (Mixture-of-Reconstruction Experts), a task-adaptive MoE framework for general-purpose MRI reconstruction. To the best of our knowledge, this is the first application of MoE to physics-constrained, end-to-end variational-network MRI reconstruction.
  \item We develop a MoE-specific training strategy for end-to-end MRI reconstruction and analyze the learned routing behavior to assess modality-aware expert specialization under different MoE configurations.
  \item We evaluate MoRE on the fastMRI datasets \cite{zbontar2018fastMRI}. Our results indicate that, with comparable per-sample computation, a properly trained conditionally sparse MoE network can better balance reconstruction performance across different distribution modes in multi-contrast datasets than a dense baseline.
\end{itemize}

\begin{figure*}[t] 
  \centering
  \includegraphics[width=\textwidth]{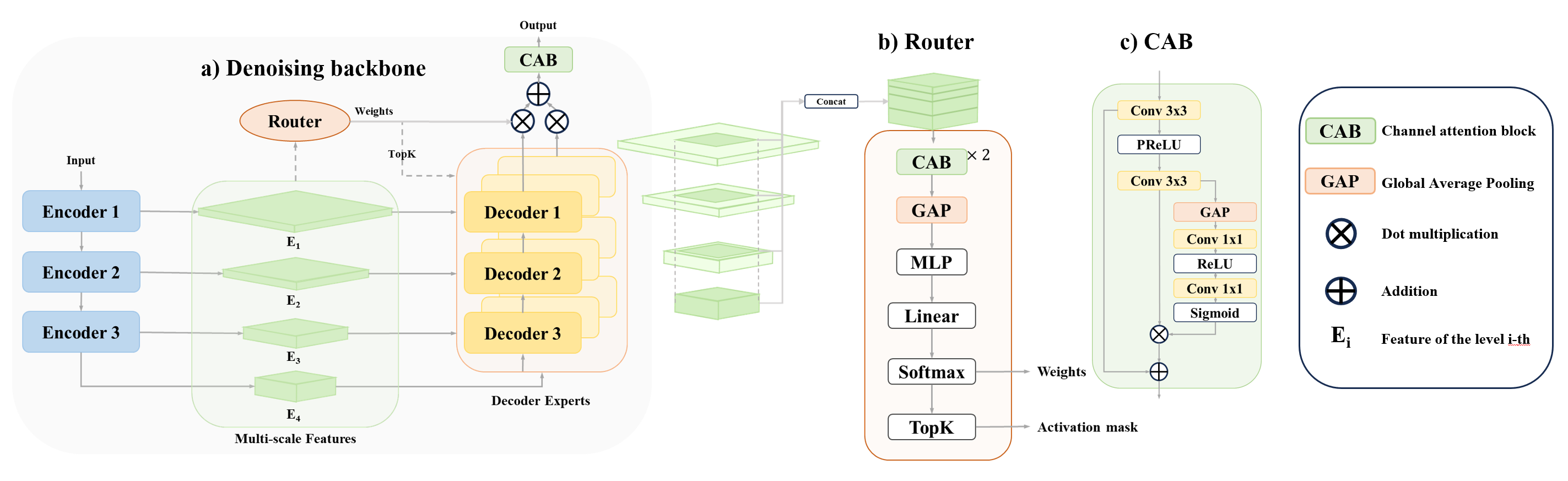} 
  \caption{The framework of our MoRE, comprising (a) a U-Net backbone with expert decoders, (b) router, and (c) CAB.}
  \label{fig:frame}
\end{figure*}

\section{Related Work}
\label{sec:relatedwork}

Considering the parallel MRI acquisition model \(y=Ax\) for an image \(x\), where \(y\) represents the measured k-space data, we estimate \(x\) by solving the following optimization problem:

\begin{equation}
    \hat{x} = \mathop{\mathrm{arg\,min}}\limits_{x} \left\{ \frac{1}{2}\,\|Ax-y\|_2^2 + \lambda\,\Psi(x) \right\}
\end{equation}

which can be solved iteratively based on E2EVarNet\cite{sriram2020endtoendvariationalnetworksaccelerated} by unrolling:

\begin{equation}
x_{c+1}=x_c-\eta_c A^\ast\left(A x_c - y\right)+\mathrm{D}(x_c)
\end{equation}

where \(\hat{x}\) denotes the estimated image. \(A\) is the forward operator that applies coil-sensitivity encoding followed by subsampled Fourier encoding. \(\Psi\) is a denoising regularizer and \(\lambda\) is its weighting coefficient. The unrolled network comprises cascades indexed by \(c\), each alternating a data-consistency step with a learned denoiser. Supervised training uses paired data consisting of simulated undersampled inputs \(x\) and fully sampled targets \(y\). \(D\) is a learnable denoising network.

\section{Method}
\label{sec:method}

In this paper, we propose MoRE, a U-Net-based denoising network \cite{ronneberger2015unetconvolutionalnetworksbiomedical} that employs a shared encoder to extract multi-scale features, an unsupervised router to adaptively classify samples from different modalities, and dynamically selects and activates different expert decoders to map them to the task space (as Fig.~\ref{fig:frame}).

\subsection{Routing and Experts Weighting}

We introduce a sample-wise router that uses multi-scale features to enable modality-adaptive, semantically meaningful expert selection, with minimal overhead and maximal reuse of existing features. Using a U-Net \cite{ronneberger2015unetconvolutionalnetworksbiomedical} with three downsampling stages as an example, all inputs to the router are detached from the computation graph so that router gradients do not perturb the high-fidelity, data-consistent features required for reconstruction, thereby stabilizing backbone convergence.

We first concatenate, along the channel dimension, features from different encoder levels $\{E_1, E_2, E_3, E_4\}$. To reduce computation while keeping fine detail, we center-crop the high-resolution features ($E_1$–$E_3$) to match $E_4$. Since the router gradients are detached from the backbone, we strengthen the router with a lightweight auxiliary two-layer convolutional feature extractor (with nonlinearity and channel attention \cite{zhang2018imagesuperresolutionusingdeep}). The scale-adaptive feature extraction is as follows:

\begin{table*}[t]
\centering
\caption{Reconstruction metrics across modalities.}
\label{tab:ssim-psnr}
\begin{tabular}{lccccccc}
\hline
& \multicolumn{7}{c}{SSIM (\%)/PSNR (dB)} \\
\cline{2-8}
Method & AXFLAIR & AXT1POST & AXT2 & AXT1 & AXT1PRE & knee-fs & knee-nofs \\
\hline
E2EVarNet & \textbf{89.80}/34.87 & 93.99/37.16 & 91.85/33.88 & \textbf{90.79}/35.84 & 92.17/35.99 & 84.14/34.53 & 92.59/35.99 \\
PromptMR-plus & 88.88/34.94 & 94.53/37.54 & \textbf{92.71}/\textbf{34.81} & 89.87/36.08 & 92.37/35.94 & 84.40/35.01 & 93.37/36.56 \\
Ours      & 89.50/\textbf{35.32} & \textbf{94.69}/\textbf{38.11} & 92.69/34.79 & 90.42/\textbf{36.33} & \textbf{92.43}/\textbf{36.42} & \textbf{84.92}/\textbf{35.34} & \textbf{93.64}/\textbf{36.96} \\
\hline
\end{tabular}

\end{table*}


\begin{figure*}[!t]
  \centering

  \includegraphics[width=\linewidth]{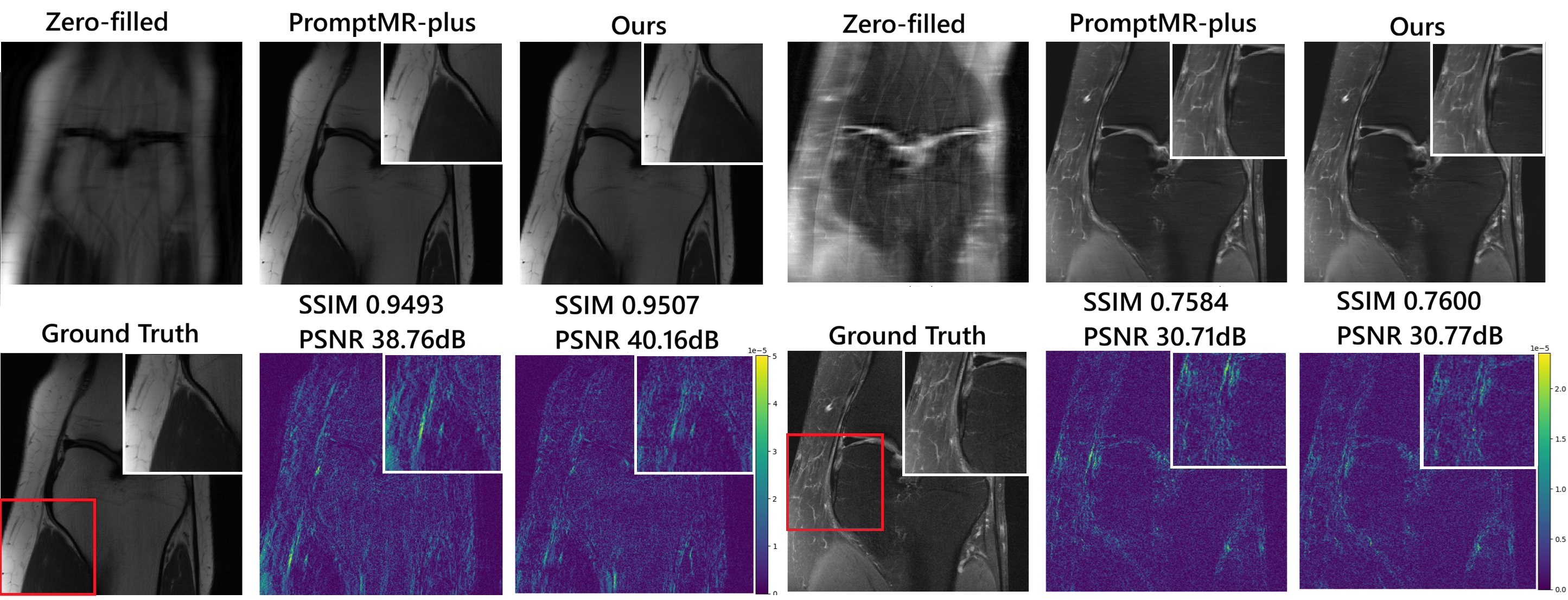}
  \\ (a) Reconstruction cases
  \vspace{6pt} 

  \begin{minipage}[t]{0.48\linewidth}
    \centering
    \includegraphics[width=\textwidth]{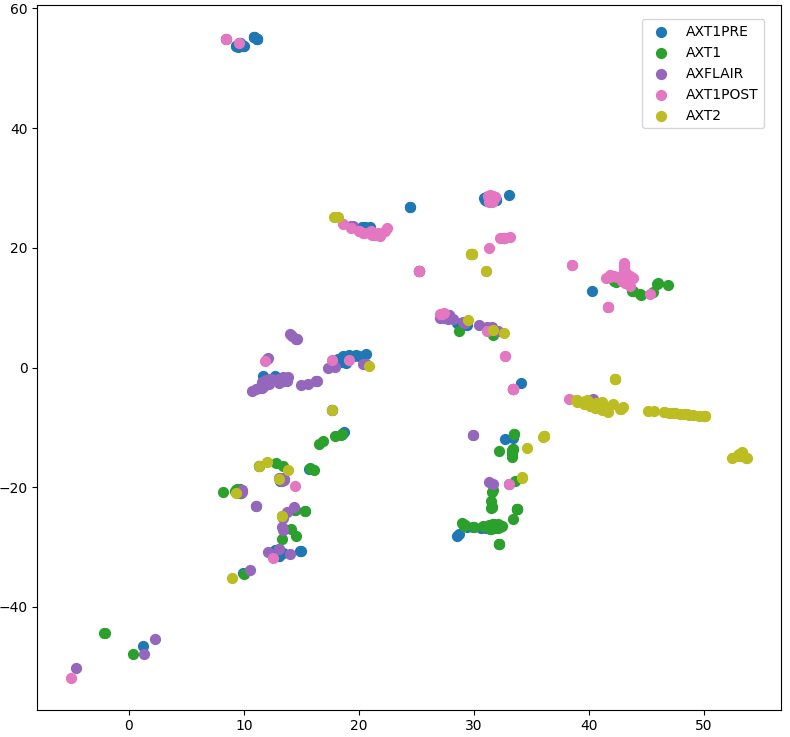}
    \\ (b) t-SNE of Brain
  \end{minipage}
  \hfill
  \begin{minipage}[t]{0.48\linewidth}
    \centering
    \includegraphics[width=\textwidth]{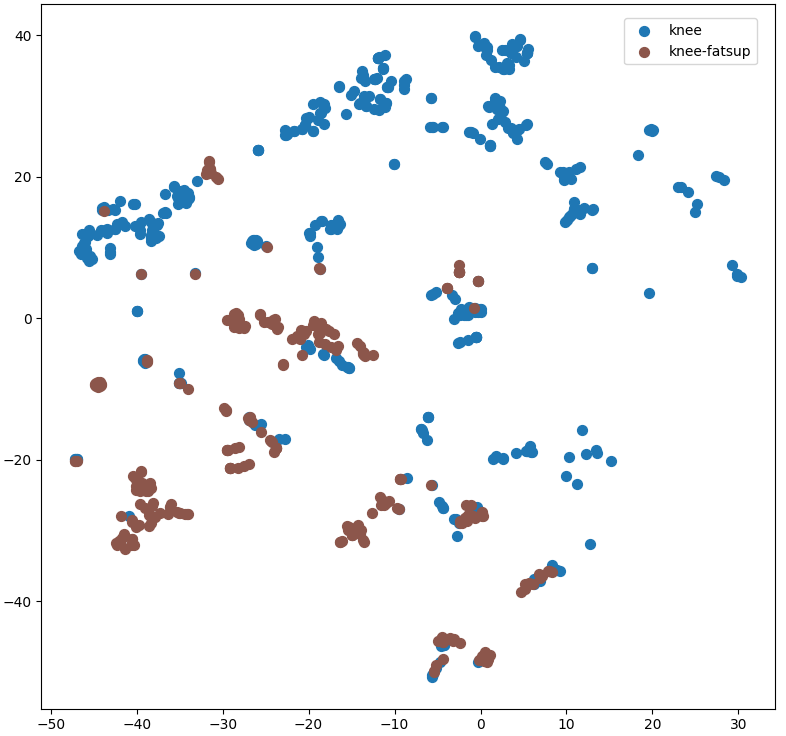}
    \\ (c) t-SNE of Knee
  \end{minipage}

  \caption{Reconstruction examples and t-SNE visualizations of the routing scheme. 
  (a) Reconstruction results for non--fat-suppressed (left) and fat-suppressed (right) knee images.
  (b--c) t-SNE visualizations for brain and knee datasets.}
  \label{fig:result}
\end{figure*}

\begin{equation}
    E_r = \mathrm{CAB}^2(\mathrm{Cat}_c(\mathrm{Crop}(\{E_i\}))), \quad i=1,2,3,4
\end{equation}

\begin{equation}
    F_r = \mathrm{MLP}(\mathrm{GAP}(E_r))
\end{equation}

where $ \mathrm{Crop}(\cdot)$ center-crops the input set to the smallest spatial size, $ \mathrm{Cat_c}(\cdot)$ denotes concatenation along the channel dimension, and $ \mathrm{CAB}^n$ denotes a stack of $n$ channel-attention convolutional blocks (as Fig.~\ref{fig:frame}c). The combination of global average pooling (GAP) and a Multilayer Perceptron (MLP) maps $E_r$ to a fixed-dimensional embedding vector.

We then use the aggregated multi-level semantic feature \(F_r\) to tailor the routing scheme \(G(\cdot)\) and to select the active decoder(s) along with their mixture weights for each sample. We use Linear Gating~\cite{mu2026comprehensivesurveymixtureofexpertsalgorithms}. A linear projection \(z = W F_r\) maps \(F_r\) to a vector \(z\) with a dimensionality equal to the number of experts. Then a set of weights \(\{g_i\}_{i=1}^{N}\) for \(N\) experts is calculated as follows:

\begin{equation}
    g_i = \frac{\exp(z_i)}{\sum_{j=1}^{N}\exp(z_j)}
\end{equation}

\begin{equation}
    G(\cdot) = \sum_{i \in \mathcal{K}} g_i\, \mathrm{Dec}_i(\cdot), \mathcal{K} := \operatorname{TopK}(\{g_i\},\,K)
\end{equation}

where \(\mathrm{Dec}_i\) denotes the \(i\)-th expert decoder, and \(\operatorname{TopK}\) returns the indices of the \(K\) largest values in the set.

\subsection{Expert Load Balancing and Training Strategies}

In sparsely activated MoE networks, the router tends to favor faster-learning experts rather than encouraging task-aligned specialization. This bias can self-reinforce and rapidly collapse into a de facto single-expert network, negating the benefits of MoE. Expert balancing aims to promote the utilization of all experts during training, enabling each to specialize on its domain within the training distribution, and thereby improving the effectiveness of the MoE. In this section, we present the load-balancing mechanism we adopt.

\subsubsection{Router Loss}

We use three losses: an auxiliary loss penalizing popular experts, a sparsity loss sharpening the weight distribution, and a z-loss bounding logits for stability, which are computed as follows:

\begin{equation}
\mathrm{loss}=
\underbrace{\alpha_{1}\sum_{i=1}^{N} p_i g_i}_{\text{aux-loss}}
\;-\;
\underbrace{\alpha_{2}\sum_{i=1}^{N} g_i \log g_i}_{\text{sparse-loss}}
\;+\;
\underbrace{\alpha_{3}\left(\log \sum_{i=1}^{N} e^{\,z_i}\right)^{2}}_{\text{z-loss}}
\end{equation}


where $p_i$ denotes the normalized probability computed from the count (rather than the weight) of samples already routed to the $i$‑th expert in the batch after the $\operatorname{TopK}$ step. The coefficients ${\alpha_i}$ are hyperparameters. Note that the auxiliary loss and the sparsity loss operate over similar scopes but encourage opposing behaviors; their weights should be tuned carefully.

\subsubsection{Noise and annealing}

We inject additive Gaussian noise $\varepsilon$, controlled by a temperature parameter $\tau$, into the TopK selection only, which is computed as follows:

\begin{equation}
\operatorname{TopK}(\{g_i\},\,K)
=
\left\{
\mathbf{1}\!\left[g_i + \tau \varepsilon_i \ge \theta_K\!\big(g + \tau \varepsilon\big)\right]
\right\}_{i=1}^N.
\end{equation}

where $\theta_K(\cdot)$ denotes the $K$-th largest threshold. We use a larger $\tau$ at the beginning of training to ensure a balanced and sufficient initialization between experts, and maintain a smaller $\tau$ in later stages to help similar samples cross boundaries, thus achieving a balance between specialization and generalization.

\subsubsection{Dummy batch}

End-to-end MRI reconstruction typically uses tiny batches (often 1), risking trivial solutions of router unsupervised clustering; we mitigate this by only augmenting each iteration’s router batch with $K$ cached input from recent iterations, improving robustness at low cost.

Note that for all router inputs, including cached batches, the gradients are detached from the backbone denoising network so that the router losses do not influence the training of the backbone network.

\section{Results}

We conducted experiments on the NYU fastMRI multi-coil brain and knee datasets \cite{zbontar2018fastMRI}. To enable controlled comparisons under a feasible experimental budget, we used a subset of 400 training volumes selected by stratified random sampling from seven fastMRI categories. Specifically, we selected 50 volumes from each of the five brain contrasts (AXT1, AXT1PRE, AXT1POST, AXT2, and AXFLAIR) and 75 volumes from each of the two knee acquisition types (fat-suppressed and non-fat-suppressed), resulting in 250 brain volumes and 150 knee volumes. The larger number of knee volumes per acquisition type was used to avoid under-representing knee data in the mixed-anatomy training set, since the knee dataset contains fewer acquisition categories than the brain dataset. We retained the original fastMRI contrast labels rather than merging visually similar categories, since these labels may reflect acquisition- or site-dependent distributional differences. All experiments used an 8$\times$ undersampling mask. We further partitioned the official fastMRI validation set into validation and test subsets with an equal number of samples per modality.

All networks are configured with 12 cascades and channel widths of [16, 20, 24, 28], and optimized using AdamW with a learning rate of $2\times 10^{-4}$ for up to 40 epochs. Baseline methods, including E2EVarNet \cite{sriram2020endtoendvariationalnetworksaccelerated} and PromptMR-plus \cite{Luo_Zhu_Zhang_Sun_2025}, are trained under identical hyperparameter settings. During our experiments, we observed that the E2EVarNet baseline without normalization layers exhibited mild fluctuations in validation metrics during the later stages of training, without a clear upward or downward trend. Consequently, early stopping with patience of 3 was applied to all experimental settings.

For MoRE, the temperature parameter $\tau$ is annealed from 1.0 to 0.01, while the auxiliary loss weight is decayed from 0.1 to 0.001. The z-loss is set to 0.001, and the sparsity regularization coefficient is fixed at 0.005. 

Table~\ref{tab:ssim-psnr} shows that MoRE maintains a well-balanced reconstruction performance across multiple modalities while achieving excellent SSIM/PSNR, although it exhibits suboptimal metrics in certain modalities. Fig.~\ref{fig:result} illustrates knee reconstructions where global error maps improve more markedly than SSIM. t-SNE of routing embeddings indicates clear modality separation despite unsupervised routing.

\section{Discussion}

As shown in Table~\ref{tab:ssim-psnr}, our method achieves the highest average PSNR across modalities and obtains the best PSNR in six out of seven modalities, with only a marginal gap of 0.02 dB on AXT2. In some cases, however, the PSNR improvement is accompanied by a slight decrease in SSIM. For example, on AXFLAIR and AXT1, our method achieves the best PSNR but a slightly lower SSIM than E2EVarNet. This suggests that reducing pixel-wise reconstruction error does not always translate into improved structural similarity. One possible explanation is the implicit averaging effect introduced by expert aggregation, which may suppress noise-like variations but can also smooth fine structures or attenuate high-frequency details.

Although our method does not always achieve the highest SSIM, it obtains the best SSIM in four out of seven modalities and ranks first or second across all modalities. When ranking second, the SSIM gaps are relatively small. These results suggest that, when properly trained, a sparsely activated MoE network can provide balanced reconstruction performance across heterogeneous modalities and improve the robustness of the E2EVarNet backbone under modality-dependent distribution shifts. This supports the motivation of using conditional expert specialization for unified MRI reconstruction across different anatomies and contrasts.

We adopt sample-level decoder routing rather than finer-grained block-wise or layer-wise routing as a practical trade-off between stability, interpretability, and efficiency. The routing decision is computed from features close to the input of the expert decoder, reducing potential mismatch between the gating signal and the representation processed by the selected experts. In contrast, applying MoE at multiple decoder blocks would require repeated routing decisions and additional expert activations across variational cascades. This would substantially increase memory consumption during training, since intermediate expert activations and routing states must be stored for backpropagation, and the cost further scales with the number of MoE blocks, selected experts, and cascades. Decoder-level routing also makes expert specialization easier to interpret, since each expert corresponds to a complete reconstruction decoder.

Despite these promising results, several limitations should be acknowledged. First, MoRE is sensitive to expert load balancing and currently relies on several regularization terms and expert-training heuristics, which increases the burden of hyperparameter tuning and prolongs training. Second, our experiments are conducted on a controlled 400-volume subset rather than the full fastMRI training set; although this design enables controlled comparisons under a feasible experimental budget, larger-scale training may further reveal the scaling behavior and potential of MoRE. Future work will investigate more robust routing networks, more efficient expert aggregation schemes, and larger-scale evaluations to improve the robustness of MoE-based reconstruction while reducing engineering and computational overhead.

\section{Conclusion}

We introduced MoRE, a sparsely activated mixture-of-experts framework for end-to-end variational MRI reconstruction. By combining model-based reconstruction with sparse conditional computation, MoRE enables expert specialization for heterogeneous MRI modalities within a unified model. On the fastMRI dataset, MoRE achieves competitive or superior SSIM/PSNR across seven modalities, including the highest average PSNR. These results suggest that sparse expert specialization is a promising direction for robust multimodal MRI reconstruction.

\section*{Acknowledgment}

This work was supported in part by the National Natural Science Foundation of China (grant number 82302295). The authors also acknowledge the support of the HPC Platform of ShanghaiTech University.

\section*{Ethics Statement}

This study used the publicly available fastMRI dataset \cite{zbontar2018fastMRI}, whose data curation was conducted under an IRB-approved study and whose metadata were de-identified before release. No new human subject experiments were conducted in this work.

\bibliographystyle{IEEEtran}
\bibliography{refs}

\end{document}